\documentclass[twocolumn,preprintnumbers]{revtex4}
\usepackage{amsthm,amsmath,amssymb}

\usepackage{graphicx}
\usepackage{algorithm}
\usepackage{algorithmic}
\usepackage{xcolor}
\usepackage[utf8]{inputenc}
\usepackage{amsmath}
\usepackage{mathtools}
\usepackage[capitalise]{cleveref}
\usepackage{listings}

\newcommand{\commentoutA}[1]{}

\newcommand{\R}{\mathbf{R}}
\renewcommand{\r}{\mathbf{r}}

\newcommand{\X}{\mathbf{X}}

\makeatletter

\begin{document}
    
    \preprint{LA-UR-21-25427}
    
    \title{Quantum-based Molecular Dynamics Simulations Using Tensor Cores}
    
    \author{Joshua Finkelstein$^{\dagger *}$, Justin S. Smith$^\dagger$, Susan M. Mniszewski$^\ddagger$, Kipton Barros$^\dagger$, Christian F. A. Negre$^{\dagger *}$, Emanuel H. Rubensson$^{\dagger \dagger}$, Anders M. N. Niklasson$^\dagger$}
    \email{jdf@lanl.gov, cnegre@lanl.gov, amn@lanl.gov}
    \affiliation{$^\dagger$Theoretical Division, Los Alamos National Laboratory, Los Alamos, New Mexico 87545}
    \affiliation{$^\ddagger$Computer, Computational, and Statistical Sciences Division, Los Alamos National Laboratory, Los Alamos, New Mexico 87545}
    \affiliation{$^{\dagger \dagger}$Division of Scientific Computing, Department of Information Technology, Uppsala University, Box 337, SE-751 05 Uppsala, Sweden}
    
    \date{\today}
    
    \begin{abstract}
        Tensor cores, along with tensor processing units, represent a new form of hardware acceleration specifically designed for deep neural network calculations in artificial intelligence applications. Tensor cores provide extraordinary computational speed and energy efficiency, but with the caveat that they were designed for tensor contractions (matrix-matrix multiplications) using only low precision floating point operations. Despite this perceived limitation, we demonstrate how tensor cores can be applied with high efficiency to the challenging and numerically sensitive problem of quantum-based Born-Oppenheimer molecular dynamics, which requires highly accurate electronic structure optimizations and conservative force evaluations. The interatomic forces are calculated on-the-fly from an electronic structure that is obtained from a generalized deep neural network, where the computational structure naturally takes advantage of the exceptional processing power of the tensor cores and allows for high performance in excess of 100 Tflops on a single Nvidia A100 GPU. Stable molecular dynamics trajectories are generated using the framework of extended Lagrangian Born-Oppenheimer molecular dynamics, which combines computational efficiency with long-term stability, even when using approximate charge relaxations and force evaluations that are limited in accuracy by the numerically noisy conditions caused by the low precision tensor core floating-point operations. A canonical ensemble simulation scheme is also presented, where the additional numerical noise in the calculated forces is absorbed into a Langevin-like dynamics. 
    \end{abstract}
    
    \keywords{electronic structure theory, molecular dynamics, 
        density functional theory,
        Born-Oppenheimer molecular dynamics}
    \maketitle

    \section{Introduction}
    
    Quantum-based Born-Oppenheimer molecular dynamics (QMD) provides a predictive and intuitively clear approach to atomistic simulations and has a broad range of application to materials science, chemistry and molecular biology \cite{RCar85,DRemler90,MCPayne92,marx_hutter_2009,BKirchner12,Negre2021-gi}. 
    For QMD simulations based on the Born-Oppenheimer approximation \cite{MBorn27,DMarx00}, the interatomic forces are calculated from a quantum-mechanical description of the electronic structure in its fully relaxed ground state as if the atoms had fixed positions in each instant of time. This is motivated by the large difference in the speed and mass between the fast, light electrons and the slow-moving, heavy nuclei. With the calculated forces we can then evolve the atoms in time as in classical Newtonian mechanics to a new configuration. The process is repeated, with new forces calculated on-the-fly for each new configuration, often over hundreds of thousands, and even millions, of time steps. The generated molecular trajectories and electronic structure from each time step provide a transparent and highly detailed picture helping us to understand and predict properties for a given material or chemical system. 
    
    In QMD simulations, the electronic structure is often calculated using density functional theory \cite{PHohenberg64,RParr89,RDreizler90} or some semi-empirical method \cite{JPople65,MDewar77,MElstner98,JStewart13,CBannwarth20}, where the electronic ground state is obtained from the solution of a non-linear eigenvalue equation. The computational cost of solving a standard linear eigenvalue equation is high and scales cubically with the number of eigenstates \cite{GGolub96}. Because of the non-linearity, however, the electronic structure can only be calculated iteratively, as in a Newton scheme, from repeated solutions to a sequence of linearized eigenvalue problems -- until a desired accuracy is met. If the solution to the non-linear eigenvalue equation is not well-converged at each time step of the simulation, the forces may not be sufficiently accurate or conservative, likely invalidating the simulation. The computational cost associated with QMD simulations therefore is often prohibitively high, limiting most applications to small systems and short simulation times. 
    
    Fortunately, with the continued rapid expansion of available computational processing power, QMD is becoming an increasingly valuable tool in both research and industrial applications. In practice, an advance in computational speed alone will frequently only have a limited effect on our ability to extend the applicability of scientific computations. Sometimes, however, it is possible to reformulate the underlying physics of a problem and recast the relevant equations in a framework that is more suitable to new solvers, algorithms, and data structures that are well-adapted to new computer architectures. In this way it is possible to take full advantage of the increase in processing power and a dramatic acceleration can often be achieved. This interdisciplinary coordinated design approach has proven to be successful throughout the history of scientific computing and is a key ingredient in the framework for QMD simulations using tensor cores that we will present here.
    
    In this article we explore how tensor cores can be used as an effective tool to accelerate QMD simulations. Tensor cores, and the closely related tensor processing units \cite{Fasi2021-vd,nvidia-tc,CYoung17,ZPan21}, are a new form of hardware designed for calculations involving deep neural networks in machine learning applications and provide an extraordinary amount of computational speed and energy efficiency \cite{Kharya2019-yy}. However, peak performance is limited to tensor contractions, i.e.\ matrix-matrix multiplications, using only low, mixed-precision floating point operations. This represents a significant obstacle to the general scientific calculations necessary for QMD simulations. QMD simulations are highly sensitive to errors in the forces, and the low precision in the tensor core floating-point operations may prevent an accurate, tight convergence of the electronic structure calculations to the iterative solution of the quantum-mechanical non-linear eigenvalue problem. Using tensor cores to accelerate QMD simulations therefore presents a particularly challenging problem. Still, based on a carefully crafted coordinated design approach, we will demonstrate how accurate QMD simulations can be performed with high efficiency using tensor cores. This opens up numerous new avenues for applications of tensor cores and tensor processing units in chemistry, materials science, and molecular biology. 
    
    The development presented in this article mirrors a similar a transition that occurred over a decade ago when graphics processing units (GPU’s) gradually became more accessible to scientific computing \cite{JStone10,TGermann09,TMartinez08,TMartinez09a,TMartinez09b,TMartinez11,JMaia12,MHacene12,FLiu2015,WHuhn20,MGordon20,ZGuoqing20,SGoedecker09}. Some of the more demanding computational tasks performed on the general purpose central processing units (CPUs) were successfully transferred to the more specialized but higher performing GPUs. Today, a broad range of GPU solvers and libraries are available using basic matrix and vector algebra that often demonstrate exceptional performance for a wide variety of general scientific applications \cite{cuSOLVER,BML,cublas}.
    
    Our article is outlined as follows. First we present some background on density functional theory, QMD, and how the electronic structure can be calculated using a novel generalized deep neural network formulation\cite{JFinkelstein21a} that naturally takes advantage of the optimized structure of the tensor cores. We then present the framework of extended Lagrangian Born-Oppenheimer molecular dynamics \cite{ANiklasson06,ANiklasson08,ANiklasson14,ANiklasson17}, which combines computational efficiency with long-term stability, even under the approximate and numerically noisy conditions caused by the low precision tensor core floating point operations. Thereafter we discuss an alternative strategy to deal with noisy data, where we view the numerical errors in the calculated forces as a natural part of a Langevin-like dynamics. These combined techniques enable a drastic increase in the computational efficiency for QMD simulations compared to state-of-the-art methods using CPUs or GPUs. We conclude with a brief summary. 

    \section{Electronic Structure Theory}

    \subsection{Density functional theory}

    In Hohenberg-Kohn density functional theory
    \cite{PHohenberg64,RParr89,RDreizler90} the ground state electron density, $\rho_{\rm min}(\r)$, is given from a constrained minimization over all $v$-representable densities, $\rho \in v$ (i.e.\ over all physically relevant densities),
    \begin{equation}\label{rho_min}
       {\displaystyle \rho_{\rm min}(\r) = \arg \min_{\rho \in v} \left\{ E(\R,\rho) \left \vert ~\int \rho(\r) d\r = N_e \right. \right\} },
    \end{equation}
    of an energy functional, 
    \begin{equation}\label{EDFT}
       {\displaystyle  E(\R, \rho ) = F[\rho] + \int \rho(\r) v_{\rm ext}(\R,\r ) d\r + V_{\rm nn } (\R)}.
    \end{equation}
    Here $F[\rho]$ is a system-independent universal functional, $v_{\rm ext}(\R,\r)$ is the external potential from the atomic nuclei, $V_{\rm nn } (\R)$ is the ion-ion nuclear repulsion, and $N_e$ is the number of electrons. 

    In Kohn-Sham (KS) density functional theory \cite{WKohn65,RParr89,RDreizler90}, which is the most common method to represent the universal functional, $F[\rho]$, the electron density is given by a sum over $N_e/2$ single-particle orbitals,
    \begin{equation}\label{charge}
    \rho({\bf r}) = 2\sum_k f_k \vert \psi_k({\bf r})\vert^2, ~~ \int \vert \psi_k({\bf r})\vert^2 d{\bf r} = 1,
    \end{equation}
    where the factor 2 is included under the assumption that each occupied orbital consists of two electrons (in spin up and down states) and the $f_k$ are the occupation factors ($f_k = 1$ for the occupied orbitals and $f_k = 0$  for the unoccupied ones). The constrained minimization of $E(\R,\rho)$, as in Eq.\ (\ref{rho_min}), using this orbital-based ansatz, can be performed by minimizing the Lagrangian function,
    \begin{equation} \label{Lagr}
        L(\rho,{\boldsymbol \epsilon}) \equiv E({\bf R},\rho) - 2\sum_k  f_k  \epsilon_k \left(\int \vert \psi_k({\bf r})\vert^2 d{\bf r} -1\right),
    \end{equation}
    with respect to the single-particle orbitals. The ground state density determined by the occupied orbitals is given from ${\delta L}/{\delta \psi_k} = 0$. Together with a KS representation of $F[\rho]$ in $E({\bf R},\rho)$, this constrained minimization condition is given in terms of the non-linear KS eigenvalue equation,
    \begin{equation}\label{KS}
        \left( -\frac{\hbar^2}{2m} \nabla^2 + V_{\rm KS}\left[ \R,\rho \right](\r)\right) \psi_k(\r) = \epsilon_k \psi_k(\r),\\
    \end{equation}
    which determines the ground state density, $\rho_{\rm min}(\r)$, by Eq.\ (\ref{charge}), from the occupied eigenstates, $\{\psi_k\}$. The Lagrange multipliers, $\epsilon_k$, and the single-particle orbitals, $\psi_k$, in Eq.\ (\ref{Lagr}), appear as eigenpairs of 
    \begin{equation}
    H_{\rm KS} \equiv  -\frac{\hbar^2}{2m} \nabla^2 + V_{\rm KS}\left[ {\bf R},\rho\right]({\bf r}),
    \end{equation} 
    the effective single-particle KS Hamiltonian obtained from the functional minimization. 
    The energy is minimized by occupying the $N_e/2$ lowest lying eigenstates of the KS Hamiltonian. The KS potential, $V_{\rm KS}\left[ {\bf R},\rho\right]({\bf r})$, depends on the electrostatic potential from the atomic nuclei at positions, ${\bf R} = \{{\bf R}_I\}$, and the electrons from the charge density, $\rho({\bf r})$, which in turn depends on the eigenstates, $\{\psi_k({\bf r})\}$. The eigenvalue equation therefore has to be solved iteratively, through the construction of a sequence of densities and KS eigenstates, 
    \begin{equation}\label{sequence}
        \rho_n \rightarrow V_{\rm KS}\left[ {\bf R},\rho_n\right] \rightarrow \{\psi_k^{(n+1)}\} \rightarrow \rho_{n+1} \ldots \rightarrow  \rho_{\rm min}
    \end{equation}
    until a converged {\em self-consistent} solution, $\rho_{\rm min}({\bf r})$, is found. This self-consistent solution is the electron density for the relaxed electronic ground state.

    \subsection{Quantum-based molecular dynamics}

    The relaxed ground state density, $\rho_{\rm min}(\r)$, determines the Born-Oppenheimer potential energy, 
    \begin{equation}
        U_{\rm BO}({\bf R}) \equiv E(\R,\rho_{\rm min}),
    \end{equation}
    which defines the QMD equations of motion,
    \begin{equation}\label{EOM}
        {\displaystyle M_I {\bf {\ddot R}}_I = - \frac{\partial U_{\rm BO}({\bf R})}{\partial {\bf R}_I}}.
    \end{equation}
    Here $M_I$ is the atomic mass at the position of the $I$-th particle with coordinates ${\bf R}_I$. These equations of motion can be integrated in time, for example, by using the time-reversible (and symplectic) Verlet scheme \cite{LVerlet67},
    \begin{equation}
        {\bf R}(t+\delta t) = 2{\bf R}(t)-{\bf R}(t-\delta t) + \delta t^2 {\bf {\ddot R}}(t).
    \end{equation} 
    
    In each new time step we must construct a new sequence of Kohn-Sham potentials to find the converged, self-consistent ground state density, $\rho_{\rm min}$ in Eq.\ (\ref{sequence}), that determines the Born-Oppenheimer potential and the forces, $-\partial U_{\rm BO}({\bf R})/ \partial {\bf R}_I$. The main cost in each force evaluation is dominated by the time it takes to find this self-consistent solution to \cref{KS} through the repeated sequence of generalized eigenvalue problems \cite{GGolub96} in \cref{sequence}. Regular eigensolvers based on diagonalization are, in general, ill suited for optimal performance on tensor cores. In our approach we avoid this shortcoming by reformulating the underlying problem such that the ground state density, $\rho_{\rm min}$, can be found through the use of a deep neural network, an ideal structure for tensor cores. To do so, we first avoid the eigenvalue problem by using the alternative density matrix representation of the electronic ground state. We then bypass the non-linearity and the iterative optimization process by using an extended Lagrangian formulation of quantum-based Born-Oppenheimer molecular dynamics. 

    \subsection{The density matrix}

    To represent the single-particle orbitals, $\{\psi_k\}$, and the density, $\rho({\bf r})$, we can use an approximate finite basis set expansion, where
    \begin{equation}
        \psi_k({\bf r}) = \sum_i^N c_i^{(k)} \varphi_i({\bf r}).
    \end{equation}
    The basis functions, $\{\varphi_i\}_{i=1}^N$, can be chosen, for example, as approximate atom-centered local atomic-orbitals. In this representation, the Kohn-Sham equation, Eq.\ (\ref{KS}), is given in terms of a generalized algebraic eigenvalue equation,
    \begin{equation}\label{KS-matrix}
        HC = SC{\boldsymbol \epsilon},
    \end{equation} 
    with the Kohn-Sham Hamiltonian matrix, 
    \begin{equation}
        H_{ij} = \int \varphi_i^*({\bf r}) \left( -\frac{\hbar^2}{2m} \nabla^2 + V_{\rm KS}[{\bf R}, \rho]({\bf r}) \right) \varphi_j({\bf r}) \; d{\bf r},
    \end{equation}
    overlap matrix,
    \begin{equation}
        S_{ij} = \int \varphi_i^*({\bf r}) \varphi_j ({\bf r})\; d{\bf r},
    \end{equation}
    and eigenvector coefficient and eigenvalue matrix,
    \begin{equation}
        C_{ij} = c_i^{(j)}, ~~ {\boldsymbol \epsilon}_{ij} = \delta_{ij}\epsilon_i.
    \end{equation}
    With this finite algebraic representation, the electron density is given by
    \begin{equation}\label{rho_D}
        \rho({\bf r}) =   \sum_{ijk} f_k c_i^{(k)}c_j^{(k)} \varphi_i^*({\bf r})\varphi_j({\bf r}).\
    \end{equation}
    However, we do not need the individual eigenstates, $\{c_i\}$, to calculate the density. Instead, we represent the electronic structure solely by the $N \times N$ density matrix,
    \begin{equation} \label{eq:DM-outer-product}
        D_{ij}  = \sum_{k} f_k  c_i^{(k)}c_j^{(k)}, 
    \end{equation}
    which contains all the necessary information about the electronic structure that is typically need in QMD simulations. The eigenvectors of the Kohn-Sham Hamiltonian, $\{c_i^{(k)}\}$, are also eigenvectors of the density matrix, but the eigenvalues of the density matrix are given by the occupation factors, $f_k$, which are 1 for states where the Hamiltonian eigenvalues, $\epsilon_k$, are below a chemical potential, $\mu$, and 0 for the states above. We can therefore calculate the density matrix directly, from a matrix step function expansion of the Kohn-Sham Hamiltonian, 
    \begin{equation}\label{theta}
        D^\perp = \theta\left(\mu I - H^\perp\right),
    \end{equation}
    which projects the occupied part of the Hamiltonian eigenvalue spectrum below $\mu$ to 1 and the unoccupied eigenvalues to 0, while the eigenvectors remain the same. 
    In this way, we can avoid the eigenvalue problem in \cref{KS-matrix}. Here, $\theta(\cdot)$ is the Heaviside step function and the step is formed at the chemical potential, $\mu$, which separates the occupied, low energy states, from the unoccupied states at higher energies ($I$ is the identity matrix). In Eq.\ (\ref{theta}) the Kohn-Sham Hamiltonian matrix, $H$, is given in an orthogonalized representation, $H^\perp = Z^T H Z$, so that the density matrix is then given by $D = ZD^\perp Z^T$ for some inverse matrix factor, $Z$, of the overlap matrix that is defined by $Z^TSZ = I$ \cite{CNegre16,GGolub96,ERubensson21}. This orthogonal representation avoids costly calculations with the overlap matrix $S$ \cite{ANiklasson05}. Once the density matrix is found, the electron density is given by $\rho(\r) = \sum_{ij} D_{ij} \varphi_i^*(\r)\varphi_j(\r)$ as in Eq.\ (\ref{rho_D}).

    \subsection{Deep-neural network electronic structure solver}

\begin{figure}
        \centering
        \includegraphics[width=.49\textwidth]{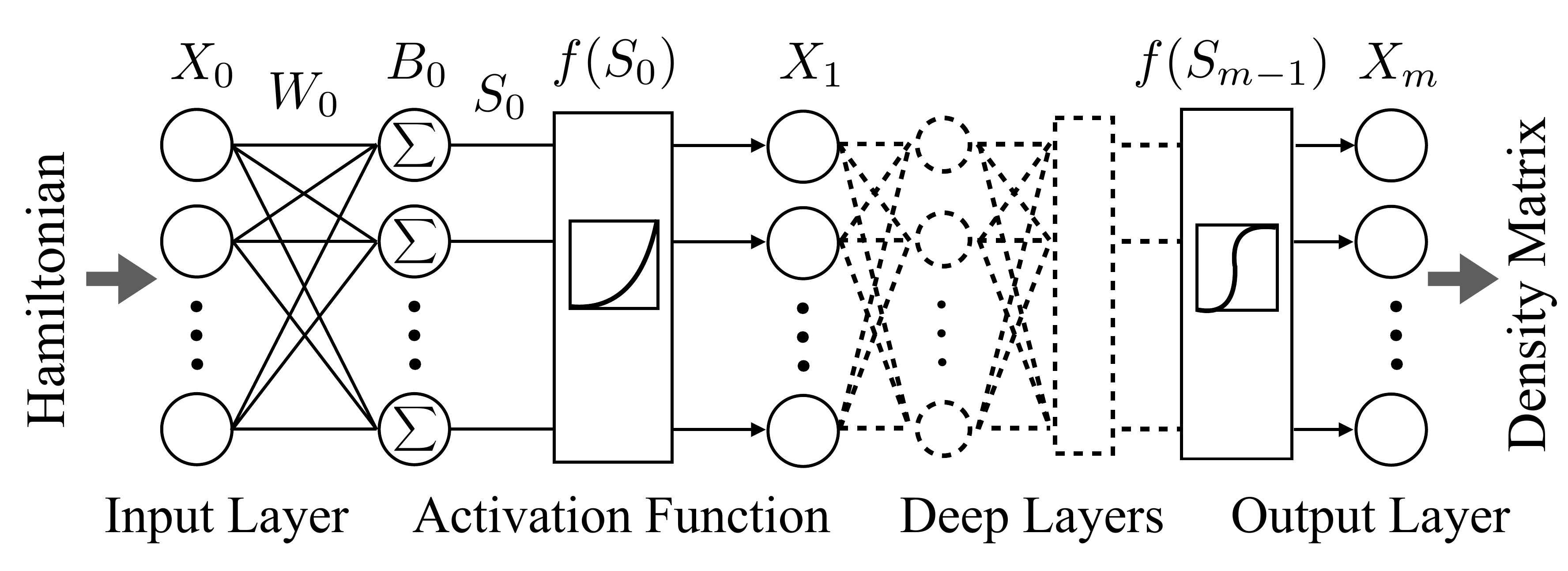}
        \caption{Schematic representing the DNN-SP2 electronic structure method for calculating the (orthogonalized) density matrix given by the matrix step function $D^\perp = \theta(\mu I-H^\perp)$, with step formed at the Fermi level, $\mu$. The input layer is given by the Hamiltonian matrix, $H^\perp$, and each deep layer $X_n$ that follows is obtained from the prior layer, $X_{n-1}$, via $X_n = f(W_nX_n+ B_n)$, where $f(X) = X^2$ is the matrix activation function, and $W_n$, $B_n$ are the weight and biases, respectively. For each $n$, $S_n = W_n X_n +B_n$ represents the $n$-th density matrix approximation so that either $S_n = {S_{n-1}}^2$ or  $S_n = 2S_{n-1} - {S_{n-1}}^2$, as in the original SP2 scheme \cite{ANiklasson02}.}
        \label{fig:Deep_NN-schematic}
    \end{figure}

    The advantage of the density matrix formulation is that we completely avoid the quantum-mechanical eigenvalue problem in Eq.\ (\ref{KS}) and instead only need to construct the density matrix by expanding the step function in Eq.\ (\ref{theta}). This expansion is achievable with a number of different techniques  \cite{RZeller82,SGoedecker93,SGoedecker94,RSilver94,APalser98,JHenk99,NBernstein01,ANiklasson02,TOzaki07,ANiklasson11,LLin13,LTruflandier16}. Several of these methods have been developed to take advantage of numerically thresholded sparse matrix algebra in the expansion of the step function, such that the computational cost only increases linearly with system size for sufficiently large sparse Hamiltonians and density matrices \cite{SGoedecker99,DBowler12}. Possibly the simplest, and most efficient, of these methods is the second-order spectral projection scheme (SP2) \cite{ANiklasson02,ERubensson11,ERubensson14}. Recently it was shown how the SP2 scheme can be written in terms of a generalized deep neural network (DNN), where
    \begin{equation}\label{Deep_NN}
        D =  f\left( \ldots f(W_1f(W_0X_0+B_0) + B_1) \ldots \right),
    \end{equation}
    and $\{W_n\}$ and $\{B_n\}$ are the weight and bias values of the neural network \cite{JFinkelstein21a}. The $n$-th density matrix approximation constructed recursively by the network is given by $S_n = W_n X_n + B_n$ with $X_n = f(S_{n-1})$. A schematic diagram illustrating the deep neural network structure, and a brief explanation, are given in Figure~\ref{fig:Deep_NN-schematic}. 
    The Hamiltonian in its orthogonal representation, $X_0 = H^\perp$,
    is used as the input parameter to the first layer. The initial weight and bias values, $W_0$ and $B_0$, are chosen to linearly transform the eigenvalue spectrum of $H^\perp$ to the interval $[0,1]$ in reverse order, i.e.\ with the lowest lying eigenvalues closer to 1 and the highest closer to 0. The activation functions, $f(X) = X^2$, are matrix square functions acting on the eigenvalue spectrum. Each layer consists of tensor contractions based on a generalized matrix-matrix multiplication that can be performed with close to peak performance on tensor cores. The computational structure of the DNN-SP2 scheme in Eq.\ (\ref{Deep_NN}) and Figure~\ref{fig:Deep_NN-schematic} therefore represents an ideal approach to performing electronic structure calculations on platforms, such as tensor cores, which have been tailored to deep neural network computations. In each layer of the DNN-SP2 network in Eq.\ (\ref{Deep_NN}), after the initial layer, the weights are chosen as $W_n = \pm I$ and biases as $B_n = (1-W_n)S_{n-1}$, with the signs selected such that the density matrix converges to the correct occupation by projecting the eigenvalues of the occupied states to 1 and the unoccupied states to 0. The layers are repeated until all eigenvalues of the output matrix are sufficiently close to 1 or 0, in which case we have an approximation of the density matrix, $D$. 
    
    The main cost in each layer of the DNN-SP2 scheme is dominated by the matrix square operation of the activation function. To account for the low precision of the tensor core floating point operations, we use a dual matrix representation,
    \begin{equation}
        X = X^{(0)} + X^{(1)},
    \end{equation}
    where, in the case of half-precision representations, the two matrices are given by
    \begin{equation}
        \begin{array}{ll}
            X^{(0)} = {\rm FP16}\left[X\right]  \\
            X^{(1)} = {\rm FP16}\left[X - X^{(0)}\right].
        \end{array}
    \end{equation}
    Here ``${\rm FP16}[X]$" denotes the half-precision (16 bit) floating point (FP) representation of $X$. The activation function can then be approximated by two separate half-precision tensor core matrix-matrix multiplications with their product accumulated in single-precision (32 bit), 
    \begin{equation}\label{activ}
        \begin{array}{l}
            f(X_n) \approx {\rm FP32}\left[ A + B + B^T \right]\\
            ~~\\
            A = X^{(0)} \times X^{(0)} ~~({\rm tensor~ core~ mult.}) \\ 
            B = X^{(0)} \times X^{(1)} ~~({\rm tensor~ core~ mult.}).
        \end{array}
    \end{equation}
    The dual mixed half-precision matrix operations double the cost of each DNN-SP2 layer, but achieves close to single-precision accuracy in the converged density matrix.  We may also add two extra layers in \cref{Deep_NN}, with a higher-order double-precision (64 bit) representation of $X$ in the activation function, which then acts as a {\em refinement} step \cite{rmcweeny56,JFinkelstein21a}, where the eigenvalues are purified to be closer to exactly 1 and 0. In this way the numerical accuracy can be further enhanced.

    \subsection{GPU and tensor core performance}
    
\begin{figure}
        \centering
        \includegraphics[width=.45\textwidth]{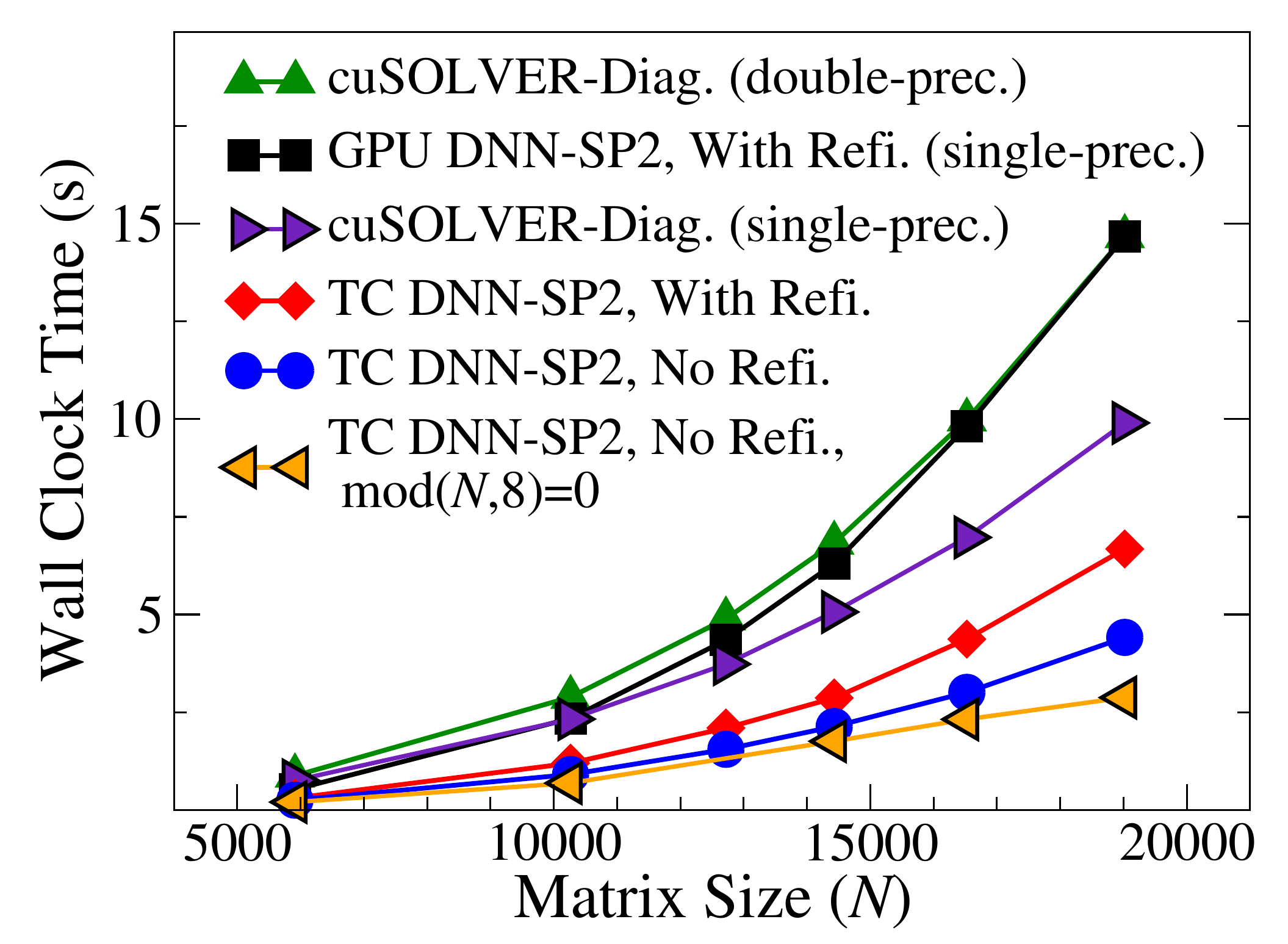}
        \caption{Wall clock time of an $N\times N$ density matrix construction for water systems with periodic boundaries using the DNN-SP2 scheme, and its variations, compared to using the cuSOLVER \cite{cuSOLVER} based single and double-precision diagonalization routines, {cusolverDnSsyevd} and {cusolverDnDsyevd}, on an Nvidia Tesla A100 GPU \cite{nvda-a100}. For $N = 19,008$, an optimal dimension for cuBLAS \cite{cublas} because ${\rm mod}(N,8) = 0$, a more than 5x speed-up is achieved with the DNN-SP2 scheme running on the tensor cores (TC) using the dual mixed half-precision matrix operations without refinement (No Refi.) when compared to the double-precision cuSOLVER-based density matrix construction. An approximate 4x speed-up is observed over the single-precision cuSOLVER routine. Between 14 and 16 deep layers were needed for convergence of the DNN-SP2 scheme.}
        \label{fig:sp2 vs cusolver}
    \end{figure}

    The generalized deep neural network structure of Eq.\ (\ref{Deep_NN}), with the activation functions calculated in mixed-precision using the dual matrix representation as in Eq.\ (\ref{activ}), provides an efficient approach to performing electronic structure calculations on tensor cores with a sequence of half-precision tensor contractions. Figure~\ref{fig:sp2 vs cusolver} shows the total wall clock time for some $N \times N$ density matrix constructions, as a function of the number of basis orbitals, $N$, for several different methods. All calculations were carried out on an Nvidia A100 GPU \cite{nvda-a100}, both with and without using its tensor cores. The input Hamiltonians were generated
    for different sizes of periodic water systems by the semi-empirical electronic structure code LATTE \cite{LATTE}, a self-consistent charge density functional tight-binding software package (see Section \ref{sec:nve}). All implementations were compiled with the PROGRESS and BML libraries \cite{BML,2016progress}, which are specialized numerical linear algebra libraries that were developed for both sparse and dense data structures when utilizing hybrid computing architectures.
    
    The first curve (upward triangles) from the top in Figure~\ref{fig:sp2 vs cusolver} shows the times for constructing the density matrix using the cuSOLVER double-precision matrix diagonalization routine cusolverDnDsyevd \cite{cuSOLVER}. This curve represents a reference state-of-the-art GPU-based electronic structure calculation for regular QMD simulations that normally would rely on double-precision floating point arithmetics. The second curve (squares) shows the times for the DNN-SP2 algorithm running on the GPU only using the cuBLAS library \cite{cublas} for matrix multiplications with single-precision floating point arithmetics. The third curve shows density matrix construction times using the single-precision cuSOLVER diagonalization, shown as (purple) frontwards triangles. The next two curves (diamonds and circles) show the timings for the DNN-SP2 scheme using the dual matrix representation and mixed half-precision tensor core calculations, again via the cuBLAS library, with and without the additional refinement (Refi.) step in the last two layers. The lowest curve (sideways triangles) displays timings for the DNN-SP2 scheme without refinement when the matrix dimensions have been slightly modified to be evenly divisible by 8 for optimal cuBLAS performance \cite{cublas}. This can be achieved either by padding $H$ with zeros or by adjusting the number of water molecules. For the DNN-SP2 schemes, these timings include both the matrix algebra as well as any auxiliary operations and so represent a combined wall clock time for all necessary calculations (e.g.\ convergence estimates, precision conversions, and memory transfers). The cuSOLVER diagonalization timings include both the diagonalization routine and the density matrix construction from the outer-product of eigenvectors as in \cref{eq:DM-outer-product}.
    
    For a matrix size of $19,020 \times 19,020$ (corresponding to over 3,000 water molecules), the complete density matrix construction with the DNN-SP2 algorithm takes about 6.6 s with the refinement step and 4.4 s without it. Due to cuBLAS design specifications, as stated above, all wall clock times for DNN-SP2 methods are substantially reduced when the size of the matrix dimensions are divisible by 8, i.e. 
    ${\rm mod}(N,8)=0$, as is presented in Figure~2. For example, when using a $19,008 \times 19,008$ matrix instead of a $19,020 \times 19,020$ (i.e.\ with 2 water molecules removed), the DNN-SP2 density matrix construction on tensor cores without the refinement step now takes only 2.87 s, a 35\% reduction in wall clock time. This is a more than 5x increase in efficiency when compared with the density matrix construction using the cuSOLVER diagonalization in double-precision, which requires 14.67 s for a matrix size of $19,020 \times 19,020$, and an approximately 4x speed-up compared to the single-precision diagonalization. The cuSOLVER GPU-based density matrix construction did not noticeably benefit from the adjustment of matrix dimensions to be evenly divisible by 8. For the remaining tensor core DNN-SP2 without refinement data shown in Figure~2, we see a reduction in times between 25\% and 30\% depending on the size of $N$ when using these favorable dimensions versus when not. A further decrease in wall clock time, by up to 50\%, can be achieved by using an accelerated version of the DNN-SP2 scheme \cite{JFinkelstein21a,ERubensson11,ERubensson14}. Though, this acceleration scheme requires prior estimates of the largest occupied eigenenergy and lowest unoccupied eigenenergy and is therefore not considered here. 

    For the tensor core implementation of the DNN-SP2 scheme, the convergence control is performed on the CPU and is given through idempotency conditions, i.e.\ the deviation from all eigenvalues being 0 or 1, which are estimated from traces of intermediate matrices \cite{JFinkelstein21a}. These matrix traces are calculated on the GPU, whereas all the time consuming generalized matrix multiplications (tensor contractions) are performed on the tensor cores. The separate dual mixed-precision tensor contractions achieve a performance of over 130 Tflops with the largest system sizes \footnote{Nvidia's advertised FLOP rate for A100 Tensor core units is approximately 300 Tflops, however in this measure additions and multiplication ($a*b + c$) are counted as two separate operations. Instead we count the combined multiplication and addition as a single floating point operation.} which is consistent with the original implementation of the DNN-SP2 algorithm in Ref.~\citenum{JFinkelstein21a}. The whole routine (not including the initial and final data transfers) has an effective peak FLOP rate of over 100 Tflops for the larger system sizes.

    In practice, the demonstrated five-fold increase in computational speed with tensor cores over the current state-of-the-art GPU-based cuSOLVER diagonalization is only achievable if we can provide stable and accurate QMD simulations with the reduced numerical precision. This is a significant challenge, because of the numerical sensitivity in the iterative non-linear optimization of the electronic ground state required prior to the force calculations. Though, by formulating QMD within the framework of an extended Lagrangian dynamics, these accuracy and stability problems can be avoided.

    \section{Extended Lagrangian Born-Oppenheimer molecular dynamics}

    In quantum-based Born-Oppenheimer molecular dynamics simulations, Eq.\ (\ref{EOM}), the interatomic forces are calculated from the relaxed self-consistent ground state density, $\rho_{\rm min}({\bf r})$, at each time step, as if the atoms were at stationary positions. The non-linear optimization required to find this self-consistent electronic ground state, in Eqs.\ (\ref{rho_min}), (\ref{KS}) and (\ref{sequence}), can be very sensitive to inaccuracies caused by numerical errors. Without a tight convergence the forces may not be sufficiently conservative, which invalidates the simulation. 
    
    \subsection{Shadow Hamiltonian approach}
    
    To avoid these expected accuracy and convergence problems, we reformulate the regular Born-Oppenheimer molecular dynamics scheme in \cref{EOM} by using a ``shadow Hamiltonian'' approach \cite{GJason00,REngel05,ShadowHamiltonian}. Instead of calculating approximate forces from an expensive iterative and numerically sensitive optimization procedure for an underlying ``exact'' Born-Oppenheimer potential, $U_{\rm BO}({\bf R})$, we can calculate exact forces directly, without the iterative procedure, but for an approximate ``shadow'' Born-Oppenheimer potential energy surface, ${\cal U}({\bf R},n)$. In this way we can avoid inaccuracies caused by the low precision floating point calculations. We achieve this by introducing an approximate ``shadow''  energy functional, ${\cal E}(\R,\rho,n) \approx E(\R,\rho)$. This approximate functional is given by a linearization of the electronic energy functional in Eq.\ (\ref{EDFT}), 
    \begin{equation}
        {\displaystyle {\cal E}(\R,\rho,n) = E(\R,n) + \int (\rho(\r) - n(\r)) \left.\frac{\delta E(\R,\rho)}{\delta \rho}\right \vert_{n}d\r} ,   
    \end{equation}
    around an approximate density, $n({\bf r})$, that is assumed to be close to the exact ground state density, i.e.\ $n({\bf r}) \approx \rho_{\rm min}({\bf r})$. The ground state optimization of this linearized functional,
    \begin{equation}\label{varrho_min}
        {\displaystyle \varrho_0[n](\r) = \arg \min_{\rho \in v} \left\{ {\cal E}(\R,\rho,n) \left \vert \int \rho(\r) d\r = N_e \right. \right\} },
    \end{equation}
    can be performed in a single step and the costly, iterative procedure, which is required for the non-linear energy functional, $E(\R,\rho)$, is avoided.
    The optimized, $n$-dependent ground state density, $\varrho_{\rm min}[n](\r)$, then provides an approximate $n$-dependent ``shadow'' Born-Oppenheimer potential, 
    \begin{equation}
        {\cal U}({\bf R},n) \equiv {\cal E}(\R,\varrho_0[n],n),
    \end{equation}
    that closely follows the exact Born-Oppenheimer potential energy surface, i.e.\ ${\cal U}({\bf R},n) \approx U_{\rm BO}({\bf R})$. The error scales as $\sim \vert \rho_{\rm min}-n\vert^2$ or $\sim \vert \varrho_{0}[n]-n\vert^2$ (e.g.\ see Ref.~\citenum{ANiklasson17}) and is therefore small as long as $n({\bf r})$ is close to the ground state. We enforce this nearness by propagating the density $n({\bf r},t) \equiv n({\bf r})$ as a time-dependent dynamical field variable through an extended harmonic oscillator that is centered around the optimized ground state. In this way, $n({\bf r},t)$ will closely follow the optimized ground state density such that the ``shadow'' Born-Oppenheimer potential energy surface is virtually indistinguishable from the regular Born-Oppenheimer potential. 
    
    \subsection{Equations of motion}
    
    The dynamics are defined by the extended Lagrangian,
    \begin{equation}
        \begin{array}{l}
            {\displaystyle  {\cal L}({\bf R,{\dot R}},n,{\dot n}) = \sum_I M_I \vert {\bf \dot R}_I\vert^2 - {\cal U}(\R,n)
            + \frac{\mu_e}{2} \int \vert {\dot n}(\r)\vert^2 d\r}\\
            {\displaystyle - \frac{\mu_e \omega^2}{2} \int (\varrho_0[n]({\bf r})- n(\r))T(\r,\r')(\varrho_0[n]({\bf r'})- n(\r')) \; d\r \; d\r'},\\
        \end{array}
    \end{equation}
    which represents a quantum-based extended Lagrangian Born-Oppenheimer molecular dynamics (XL-BOMD) \cite{ANiklasson08,ANiklasson14,ANiklasson17}.
    Here $T(\r,\r')= \int K(\r,\r'')K(\r',\r'') d\r''$ is a metric tensor for the extended harmonic well that is determined by some kernel $K(r,r')$;  $\omega$ is the frequency of the extended harmonic oscillator; and $\mu_e$ is a fictitious electronic mass parameter.  We define the kernel to be the inverse of the Jacobian of the residual function, $f[n]({\bf r}) = (\varrho_0[n]({\bf r}) - n({\bf r}))$. The equations of motion for this extended Lagrangian formulation of QMD are then derived in an adiabatic limit (similar to the Born-Oppenheimer approximation) \cite{ANiklasson08,ANiklasson14,ANiklasson17,ANiklasson20} of the Euler-Lagrange equations, as $\omega \rightarrow \infty$ and $\mu_e \omega = {\rm constant}$, which provides for a partial decoupling between the nuclear and the electronic degrees of freedom, where
    \begin{equation}\label{XL-BOMD}
        \begin{array}{l}
            {\displaystyle M_I {\bf \ddot \R}_I = \left. -\frac{\partial {\cal U}({\bf R},n)}{\partial {\bf R}_I}\right \vert_n},\\
            ~~\\
            {\displaystyle {\ddot n}({\bf r}) = - \omega^2 \int K({\bf r,r'})\left(\varrho_0[n]({\bf r'}) - n({\bf r'})\right) d{\bf r'}}.
        \end{array}
    \end{equation}
    The kernel, $K(\r,\r')$, whose square forms the metric tensor of the harmonic potential, acts like a preconditioner for the electronic equations of motion. The action of the kernel on the residual, $f[n]({\bf r'})$, can be approximated, for example, by using a preconditioned Krylov subspace approximation \cite{ANiklasson20}. A variety of integration techniques for the electronic degrees of freedom can be used \cite{ANiklasson09,PSteneteg10,GZheng11,AOdell09,AOdell11,VVitale17,ILeven19}, though for the QMD simulations in the examples below we will use the modified Verlet integration scheme presented in Ref.~\citenum{ANiklasson09}. The equations of motion for the nuclear positions include the forces given by the gradient of the shadow potential under constant density $n$, since $n({\bf r})$ occurs as a dynamical field variable. 
    
    XL-BOMD provides a general and theoretically flexible framework that can be applied to a broad range of problems with different representations of the electronic structure, including polarizable force fields and charge equilibration models \cite{KNomura15,AAlbaugh15,ANiklasson21}. XL-BOMD has many similarities with extended Lagrangian Car-Parrinello molecular dynamics \cite{RCar85,DRemler90,GPastore91,BHartke92,FBornemann98,DMarx00,HSchlegel01,SIyengar01,MTuckerman02,JHerbert04,JHutter12,JLi16}, but is based on a different form of the Lagrangian.

    The adiabatic equations of motion for XL-BOMD in Eq.\ (\ref{XL-BOMD}) provides a numerically robust approach to QMD simulations. It was shown how it even generates stable molecular trajectories with good long-term energy conservation under noisy conditions, where the forces are calculated using numerically thresholded sparse matrix algebra to reach a reduced (linear scaling) complexity as a function of system size \cite{MCawkwell12,MArita14,ANiklasson16,TOtsuka16,THirakawa17}. As we will demonstrate below, the same stability is also seen in combination with the numerically noisy tensor core calculations. In addition to a reduced numerical sensitivity, XL-BOMD provides a significant speed-up compared to regular QMD, Eq.\ (\ref{EOM}), because the costly iterative non-linear ground state optimization is avoided. 

    \subsection{Microcanonical molecular dynamics using tensor cores} \label{sec:nve}
    
    To demonstrate the efficiency and accuracy of QMD simulations using tensor cores, within the framework of XL-BOMD, where the interatomic forces are calculated from the DNN-SP2 electronic structure solver using dual  mixed-precision matrix operations, we will use approximate second-order self-consistent charge density functional tight-binding (SCC-DFTB) theory \cite{WHarrison80,MFoulkes89,DPorezag95,MFinnis98,TFrauenheim00,PKoskinen09,MGaus11,BAradi15,BHourahine20,MElstner98}. SCC-DFTB theory is based on a second-order expansion in the charge density fluctuations of the electronic energy functional in Kohn-Sham density functional theory around a reference density of overlapping neutral atomic charge distributions. The electrostatic energy is approximated by the Coulombic interactions between atomic net Mulliken charges, which is screened at short range to the chemical hardness (or Hubbard-U) terms for the on-site interactions. The fluctuating net Mulliken charges, ${\bf q} \in \mathbb{R}^{N_a}$, where $N_a$ is the number of atoms, are optimized self-consistently from the solution of a non-linear Kohn-Sham eigenvalue equation as in Eq.\ (\ref{KS}). The relaxed self-consistent ground state solution, ${\bf q}_{\rm min}$, then defines the Born-Oppenheimer potential energy surface and the forces as in Eq.\ (\ref{EOM}). In the framework of XL-BOMD we construct the shadow Born-Oppenheimer potential from the minimization of an electronic energy functional that has been linearized around an approximate solution, ${\bf n} \in \mathbb{R}^{N_a}$, to the self-consistent net Mulliken charges, ${\bf q}_{\rm min}$. This approximate charge vector (one net charge per atom) is then included as a time-dependent dynamical variable vector, ${\bf n}(t)$, that is propagated through the extended harmonic oscillator as in Eq.\ (\ref{XL-BOMD}). The main difference in SCC-DFTB theory compared to regular Kohn-Sham density functional theory thus is that the dynamical electronic degrees of freedom, ${\bf n}(t)$, is a coarse-grained net Mulliken charge vector instead of a continuous density field $n({\bf r},t)$ so that the corresponding kernel becomes an $N_a \times N_a$ matrix, i.e.\ ${\bf K} \in \mathbb{R}^{N_a \times N_a}$. This changes the equations of motion in Eq.\ (\ref{XL-BOMD}) to
    \begin{equation}\label{XL-BOMD-DTFB}
        \begin{array}{l}
            {\displaystyle M_I {\bf \ddot \R}_I = \left. -\frac{\partial {\cal U}({\bf R},{\bf n})}{\partial {\bf R}_I}\right \vert_{\bf n}},\\
            ~~\\
            {\displaystyle {\bf {\ddot n}}({\bf r}) = - \omega^2 {\bf K} \left({\bf q}_0[{\bf n}] - {\bf n} \right) },
        \end{array}
    \end{equation}
    where ${\bf q}_0[{\bf n}]$ is the optimized vector of ground state net Mulliken charges of the linearized energy functional corresponding to $\varrho_0[n]({\bf r})$ in Eq.~(\ref{varrho_min}). The integration of the density ${\bf n}(t)$ in Eq.\ (\ref{XL-BOMD-DTFB}) is performed with a modified Verlet scheme that includes a weak dissipation term in order to remove the accumulation of numerical noise and helps synchronize ${\bf n}(t)$ with the nuclear motion \cite{ANiklasson09,PSteneteg10,GZheng11}. For the nuclear coordinates and velocities we use a regular leapfrog velocity-Verlet scheme \cite{ANiklasson17}.

    \begin{figure}
        \centering
        \includegraphics[width=.49\textwidth]{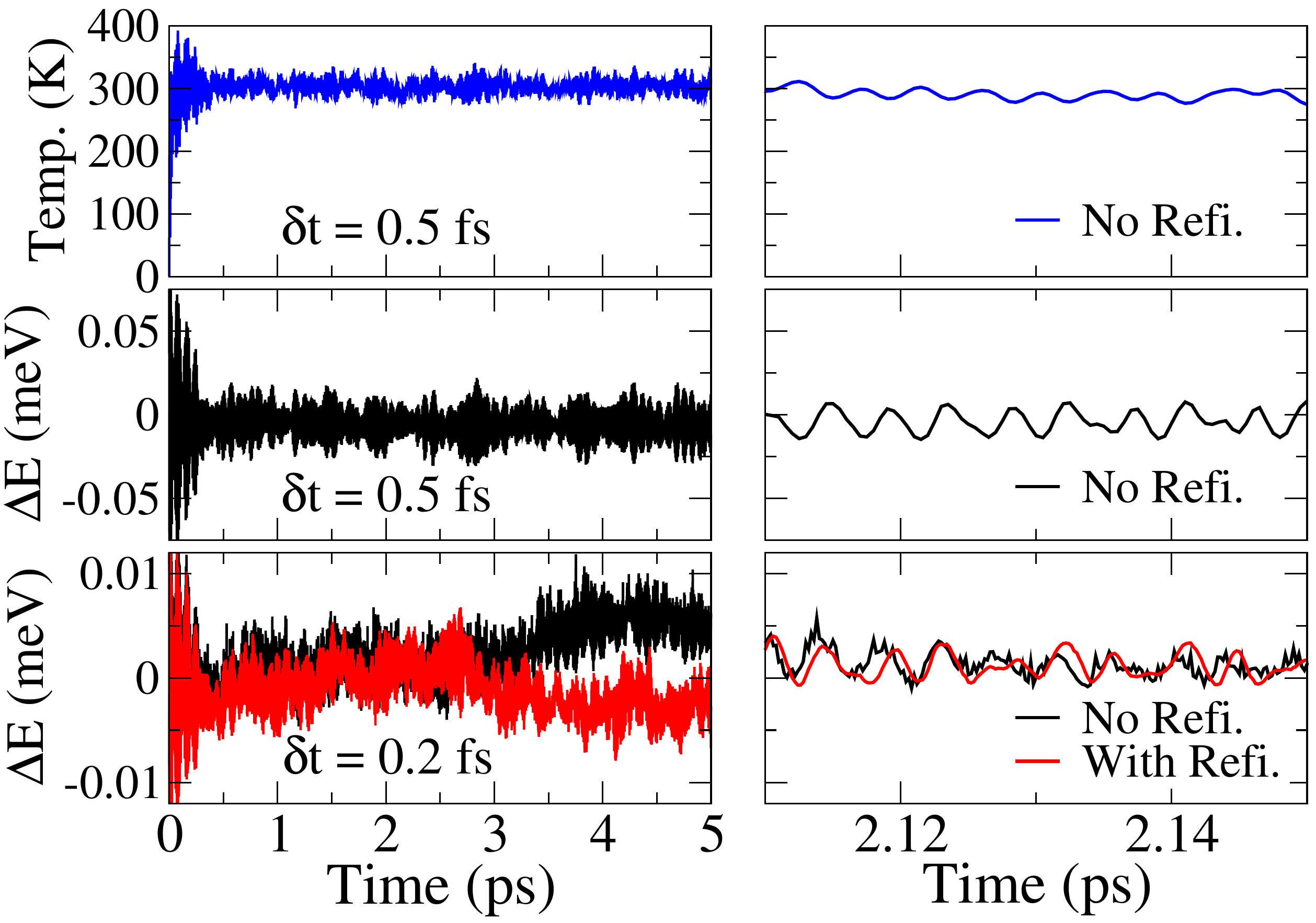}
        \caption{XL-BOMD based QMD simulations of a water system using tensor cores. The top two rows show the fluctuations in the statistical temperature (Temp.) and the total energy, $\Delta E$, for an integration time step $\delta t = 0.5$ fs.  The bottom row shows the same simulation but with a shorter integration time step, $\delta t = 0.2$ fs, including an additional refinement (Refi.) step in the last two layers of the Deep-NN SP2 algorithm (in red). The panels on the right-hand side show a shorter time snapshot of the same simulations.}
        \label{fig:WaterSimulation}
    \end{figure}
      
    Figure \ref{fig:WaterSimulation} shows the results from microcanonical XL-BOMD based QMD simulations using tensor cores for a water system with 88 molecules in a periodic box. An approximate constant kernel, $K_0 \approx K$, was used, which was calculated only once (at the first time step) and was then kept fixed during each simulation. No iterative self-consistent field optimization was used except for the very first MD time step. This alone provides a significant acceleration to the QMD simulations, because only a single density matrix construction per time step is needed. This is in contrast to regular direct BOMD which requires a tightly converged iterative sequence of density constructions in \cref{sequence} prior to the force evaluations in each integration time step. The simulations are initiated with zero atomic velocities from identical initial conditions. The panels in the middle row of Figure  \ref{fig:WaterSimulation} demonstrate stable dynamics, using an integration time step, $\delta t$ = 0.5 fs, where the fluctuations in the total energy, $\Delta E(t)$, remain stable with no visible drift caused by the noise from the limited numerical precision of the tensor cores. The panels in the bottom row of the figure show the same simulation, but with a smaller integration time step, $\delta t$ = 0.2 fs. Normally we would chose the largest possible integration time step. Here, $\delta t$ = 0.2 fs is used only to be able to detect the effects of the limited numerical precision. The amplitude of the total energy fluctuations scales as $\delta t^2$ and is therefore over 6 times smaller in the bottom row.  Only in this case, when $\delta t$ = 0.2 fs, can the effect of the numerical noise from the tensor cores be seen. A slight random-walk-like behavior in the trajectories is also seen on the left hand-side (noticed most clearly in the long-term deviation between the red and black curves); the right hand side shows a zoomed in time snapshot where we can now see a noticeable effect from the numerical noise in the total energy of the black curve. This noise vanishes for the red curve when we use an additional double-precision refinement step (With Refi.) in the last two layers of the DNN-SP2 algorithm. In the middle row ($\delta t = 0.5$ fs) the noise level is too small to be seen amongst the much larger total energy fluctuations. The behavior is very similar to what has been observed for XL-BOMD simulations using reduced complexity linear scaling solvers, for example, based on numerically thresholded sparse matrix algebra \cite{MCawkwell12,MArita14,TOtsuka16,ANiklasson16}. The examples in Figure \ref{fig:WaterSimulation} highlight a key result of this work: the ability to perform efficient and accurate QMD simulations using tensor cores. 
      
    \section{Canonical QMD Simulations using tensor cores}

    The tensor core calculations of the density matrix introduces a small, but noticeable, level of noise into the force evaluations at each MD time step, which can be observed in the total energy fluctuations in the lower panel of Figure~\ref{fig:WaterSimulation}. This noise appears similar to the noise generated by Langevin thermostats used for canonical QMD simulations. It is therefore plausible to try and reformulate the XL-BOMD scheme in \cref{XL-BOMD} into a Langevin-like dynamics where the excess noise can be damped out of the system in a carefully balanced way. Formulated as a continuous dynamics, the equations of motion in \cref{XL-BOMD} with the noisy forces and a dissipation term would become,
    \begin{align}
    \begin{split}
        d\R_I &=  {\bf \dot{R}}_I \; dt \;, \\
        d {\bf \dot{R}}_I &= \left. -\frac{1}{M_I} \bigg( \frac{\partial {\cal U}(\R , n)}{\partial \R_I}\right \vert_n + \xi_t^I + \gamma_I {\bf \dot{R}}_I \bigg) \; dt,  \\
        \ddot{n}(\r) & = -\omega^2 \int K(\r,\r') (\rho_0[n](\r')-n(\r')) \; d\r', 
    \end{split} \label{eq:noisy force}   
    \end{align}
    with a stochastic noise process $\xi_t^I$ such that  
    $\langle \xi_t^I \xi_t^J \rangle = \sigma_{_{\textrm{TC}}}^2 \delta_{IJ}$ for some unknown value $\sigma_{_\textrm{TC}}$ that is balanced by a dissipative damping term with a coefficient $\gamma_I$. The size of the error fluctuations is given by the constant $\sigma_{_\textrm{TC}}$, which in general will be system dependent and can be estimated from a comparison to double-precision calculations of the forces in the initial stages of an MD simulation (e.g. see Figure~\ref{fig:force-distribution}). 
    
    \subsection{Non-Gaussian character of tensor core noise}
    
\begin{figure}
        \includegraphics[width=.455\textwidth]{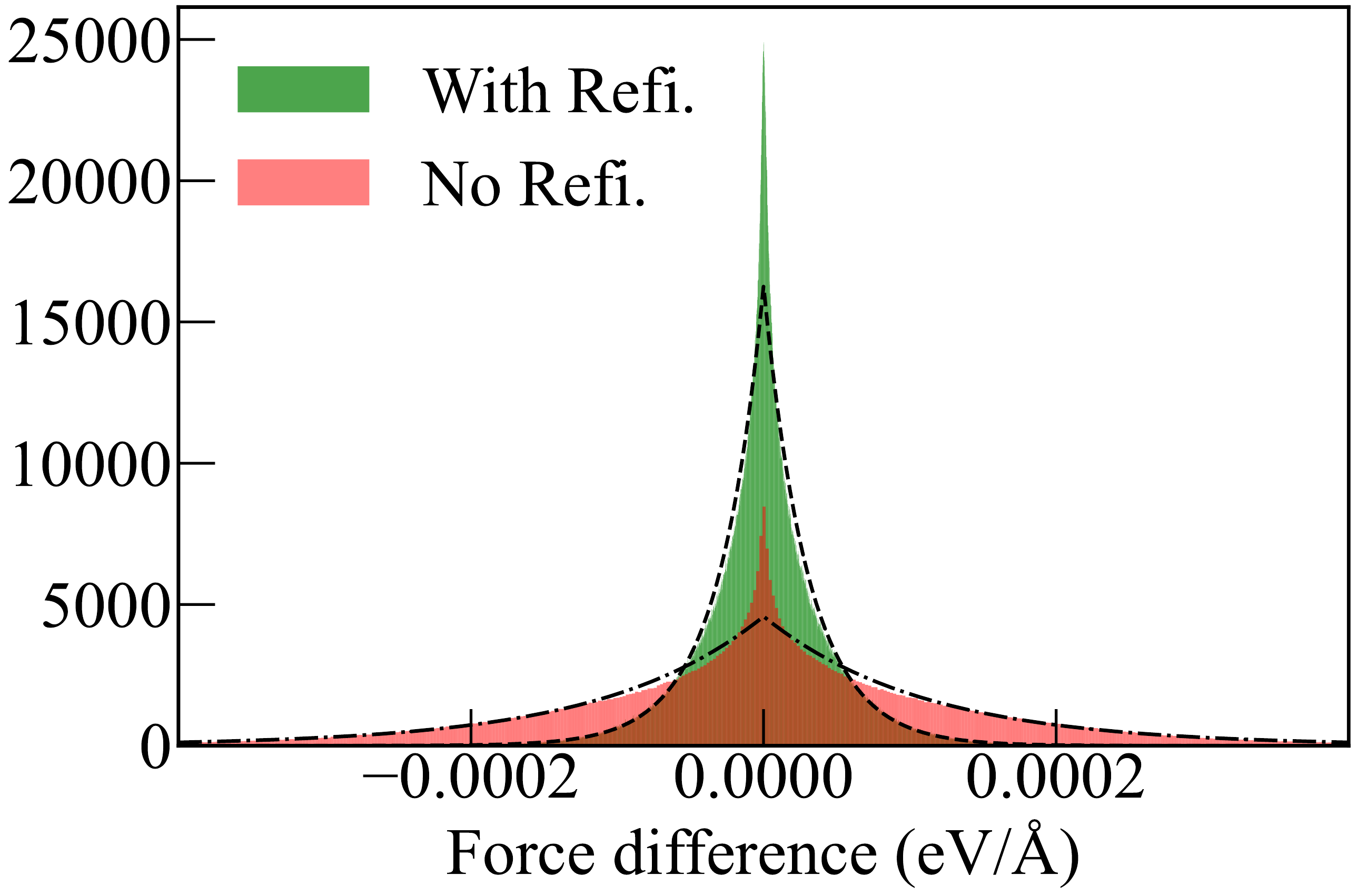}
        \caption{Force noise distribution in the $x$-coordinate, both with and without the refinement step. The dashed lines represent a Laplace distribution with the same mean and variance as given by the corresponding histogram.
        } \label{fig:force-distribution}
    \end{figure}

    The continuous dynamics in \cref{eq:noisy force} is an idealization. In practice, the noise is embedded in the force and the discretization of time induces a dynamics where the noise is integrated together with the force, simultaneously and in the same way. The numerically integrated noise term over a time interval $\delta t$ then becomes
    \begin{align}
        \sigma_{_\textrm{TC}}\delta \X_I \delta t \approx \int_{t_n}^{t_{n+1}} \xi_t^I \; dt.
    \end{align}
    Consequently, and in contrast to regular Langevin dynamics, the integrated discrete noise scales linearly with the time step, $\delta t = t_{n+1}-t_n$; and $\delta \X_I$, is in general represented by a random variable that is non-Gaussian. This cannot occur if the stochastic noise process $\xi_t^I$ is identical and uncorrelated over time (with zero mean and finite variance) as is assumed in a regular Langevin dynamics, for otherwise, the Central Limit Theorem would imply that $\delta \X_I$ is normally distributed. Instead, based on extensive numerical testing, we find that $\delta \X_I$ follows an approximate Laplace distribution with zero mean and finite variance, $\sigma_{_{\textrm{TC}}}^2$, and is independent of the size of the integration time step, $\delta t$. The Laplace-like (approximately double-sided exponential) distribution of the force errors can be estimated by a comparison between the tensor core calculations to regular double-precision calculations at each time step and is illustrated by the histogram in Figure~\ref{fig:force-distribution} for the force in the $x$-coordinate. This histogram was generated for an 88 molecule water system. The errors occur from how the eigenvalues converge to 0 and 1 during the recursive projections in the DNN-SP2 scheme, and their deviation from 0 and 1 are the main source of force error. If we use an additional double-precision refinement step, these idempotency errors are drastically reduced as can be seen in the green histogram in Figure~\ref{fig:force-distribution} and by the absence of any visible noise in the total energy fluctuations in the lower right panel of Figure~\ref{fig:WaterSimulation}. 
    
    In principle, the non-Gaussian distribution of the integrated noise terms violates the conditions of a Langevin dynamics. Although, it has been argued that if the noise term in a regular Langevin dynamics is replaced with a non-Gaussian noise of zero mean and finite variance, the same canonical averages as in a Langevin dynamics can still be recovered \cite{BDunweg1991,ALadd2009,SMelchionna07,IVattulainen2002}. Numerical examples of this phenomenon are presented in the Supporting Information document. These observations help to motivate the use of a Langevin-based model to describe the molecular dynamics trajectories with tensor core noise, and in fact, any mixed-precision noise in the force calculations, as long as the noise in the forces satisfies the conditions for the Central Limit Theorem. In effect, we therefore treat the noise from the tensor core units as a heat bath, except that now for the thermal fluctuations we use a more exotic noise of non-Gaussian character.
    
    \subsection{Canonical integration scheme}
    
    Because the integrated noise term, $\sigma_{\rm TC} \delta \X_I \delta t$, scales linearly with the integration time step, $\delta t$, in contrast to a Langevin dynamics, the regular fluctuation dissipation balance that determines the size of dissipation coefficient $\gamma$ needs to be modified in the design of an efficient and accurate integration scheme for Eq.\ (\ref{eq:noisy force}). We chose to construct a generalized integration algorithm that also includes an external regular Langevin term with a normally distributed noise term, $\eta_n$, with zero mean and unit variance, i.e.\ $\eta_n \in {\cal N}(0,1)$, and dissipation coefficient $\gamma_L$. This improves the flexibility for simulations at higher temperatures and frictions due to the fact that the $\gamma_{\rm TC}$ term is often quite small, causing slow equilibration on its own. Our algorithm, described for a single nuclear (scalar) coordinate of particle $I$ with initial velocity $V_0 = \dot{R}_{I_{\alpha}}(t_0)$, mass $M$, initial position $R_{0} = R_{I_{\alpha}}(t_0)$ and time step $\delta t$ at time $t_{k} = k \delta t + t_0$, is given by
    \begin{align}\label{IntegrationAlg}
        \begin{split}
        {V}_{k+1/4} &= {V}_k + \frac{\delta t}{2M} \left({ F}_k^{\rm TC} - {\gamma}_{\rm TC}{V}_k \right)\\
        {R}_{k+1/2} &= {R}_k + \frac{\delta t}{2} {V}_{k+1/4}\\
        {V}_{k+3/4} &= c_L \; {V}_{k+1/4} + {\sigma}_L \; {\eta}_k\\
        {R}_{k+1} &= {R}_{k + 1/2} + \frac{\delta t}{2} {V}_{k+3/4}\\
        {V}_{k+1} &=  V_{k+3/4} + \frac{\delta t}{2M} \left( F_{k+1}^{\rm TC} - \gamma_{\rm TC} {V}_{k+1} \right)\;,
        \end{split}
    \end{align}
    with $c_L = ({1-\gamma_L \delta t/2})/({1+\gamma_L \delta t/2})$ and where the force terms, $F_k^{\rm TC}$, including the embedded tensor core noise, are given by
    \begin{align}
        &{F}^{\rm TC}_k = \left. -\frac{\partial {\cal U}({\bf R}_k,n)}{\partial { R}_{I_\alpha}} \right\vert_n + { \sigma}_{\rm TC} {\xi}_k, ~~ \xi_k \in {\cal L}^{\rm TC}(0,1),
    \end{align}
    for atom $I$ and $\alpha$ an $x,y,~{\rm or}~ z$ coordinate. The relevant discrete-time fluctuation dissipation-relations are
    \begin{align} 
        & \sigma_L = \sqrt{k_{\rm B} T (1-c_L^2) / M }\;, \\
        &\sigma_{\rm TC} = \sqrt{ 2 k_{\rm B}T \gamma_{\rm TC} /\delta t}. \label{eq:fluct_diss}
    \end{align}
    These relations determine the $\gamma$-parameters for a given size of the fluctuations, $\sigma$, statistical temperature, $T$, integration time step, $\delta t$, and nuclear mass, $M$. The random noise term, $\xi_k$, in the force is assumed to have some kind of non-normal distribution, ${\cal L}^{\rm TC}(0,1)$, with zero mean and unit variance. The electronic degrees of freedom in \cref{eq:noisy force} can be integrated separately using the modified Verlet integration algorithm in Ref.~\citenum{ANiklasson09}. 
    
    When $\gamma_L=0$, we recover an integration scheme similar to the algorithm by Br\"{u}nger, Brooks and Karplus (BBK)  \cite{MKarplus1984,BLeimkuhler2016}, but for non-normal tensor core noise. The BBK method is known to require the condition $\gamma_{\rm TC} \delta t \ll 1$ \cite{RPastor88,JFinkelstein20a}. The estimated magnitudes of the tensor core noise in this work easily satisfy this condition for any relevant integration time step. A key feature of \cref{IntegrationAlg} is that we only require a single force evaluation, and therefore, a single random variable per time step. In the original BBK method, 
    \emph{two} independent random random variables are needed, thereby requiring, two separate force evaluations per time step due to the fact that randomness can only enter through a tensor core force evaluation. If we were to implement the original BBK method, it becomes necessary to modify the discrete-time fluctuation-dissipation balance in \cref{eq:fluct_diss} to 
    \begin{align}\label{eq:fluct_diss_new}
        \sigma_{\rm TC} = \sqrt{4 k_{\rm B} T \gamma / \delta t} \;.
    \end{align}
    A Python implementation of \cref{IntegrationAlg} with the fluctuation-dissipation relation in \cref{eq:fluct_diss} is presented in the Supporting Information. A brief discussion on such modifications of the BBK method is given on pg. 273 of Ref.~\citenum{BLeimkuhler2016}.
    
    For $\gamma_{\rm TC} = 0$, the algorithm in Eq.\ (\ref{IntegrationAlg}), with the given choice of $c_L$, is a splitting formulation of the method by Gr{\o}nbech-Jensen and Farago \cite{NGronbechjensen13,EMartinez15}. In fact, any of the recently derived one-parameter family of thermodynamically exact Gr{\o}nbech-Jensen (GJ) methods \cite{NGronbechjensen20,JFinkelstein20a,JFinkelstein21b} can be written using a similar splitting when an additional time-scaling factor is included. A detailed derivation and theoretical evaluation for the family of GJ integration schemes in splitting form, on which Eq.\ (\ref{IntegrationAlg}) is based, is presented in Ref.~\citenum{JFinkelstein21b}. An alternative integration method, again based on this splitting formalism, is given in the Supporting Information.
    
    \subsection{Constant temperature molecular dynamics using tensor cores} \label{sec:canonical}

\begin{figure}
        \includegraphics[width=.455\textwidth]{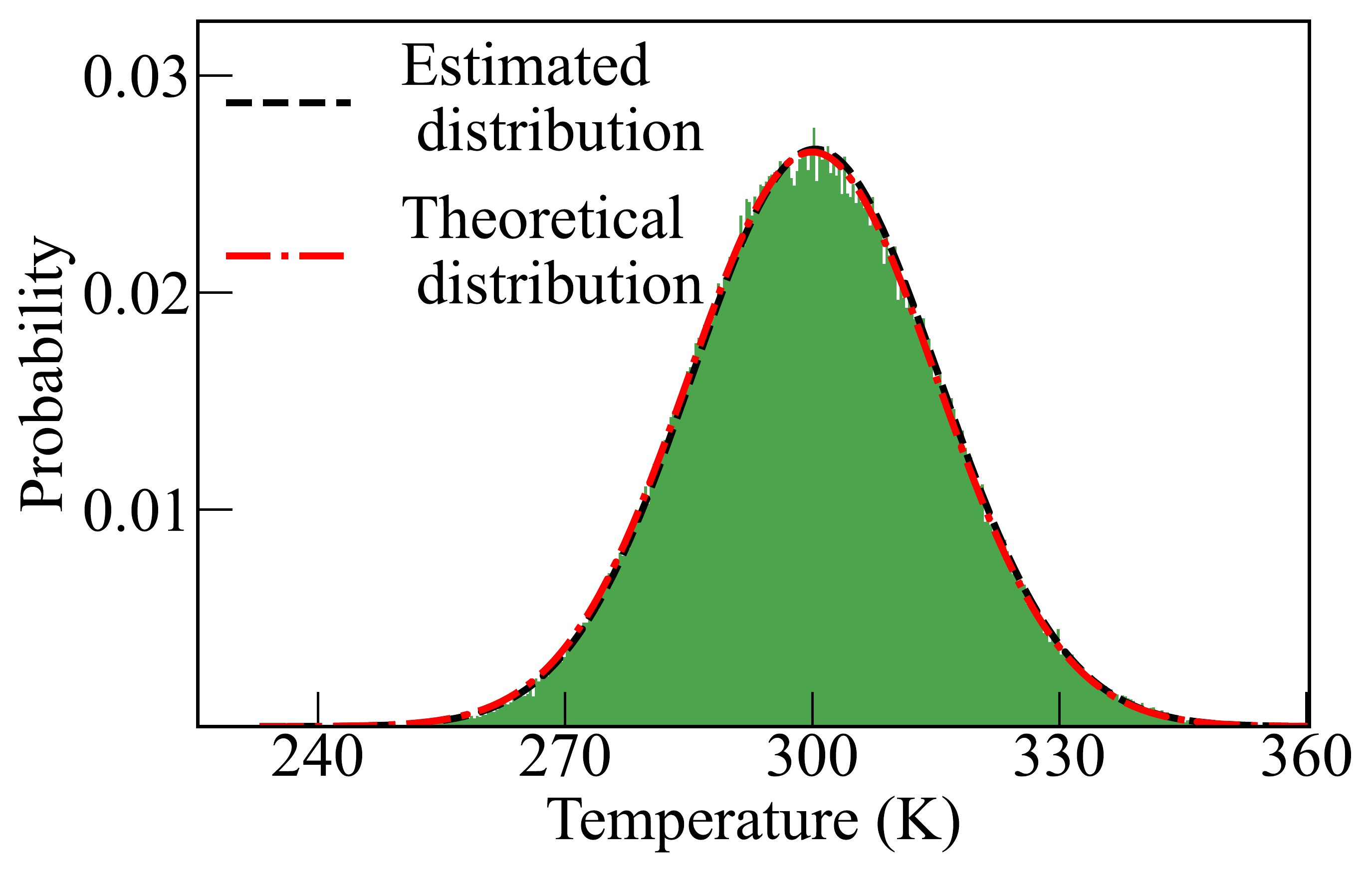}
        \caption{Statistical temperature histogram from a 100 ps canonical QMD simulation of an 88 molecule water system using tensor cores and the DNN-SP2 scheme without refinement. The integration algorithm in Eq.~(\ref{IntegrationAlg}) is used along with the regular Langevin friction term, $\gamma_L = 10^{-3}$ fs$^{-1}$, and time step $\delta t=0.5$ fs. Here, $\langle T\rangle=300$ K and $\sigma=15.07$ K are the exact theoretical values \cite{EMartinez15}. The estimated values from the simulation are $\widehat{T}=300.37$ and $\widehat{\sigma}=14.99$ K. Temperatures are calculated using the half-step velocity in Ref.~\citenum{NGronbechjensen19}. } \label{fig:temp-distribution}
    \end{figure}

    The equations of motion in Eq.~(\ref{eq:noisy force}), including additional Gaussian noise, can be integrated using a combination of the integration schemes for the electronic degrees of freedom described in Ref.~\citenum{ANiklasson09} and our proposed algorithm in \cref{IntegrationAlg}. To illustrate this method for a canonical (NVT) simulation, we use an 88 molecule water system in a periodic box. The QMD simulations were performed within the framework of XL-BOMD and based on SCC-DFTB theory where the density matrix is calculated with the DNN-SP2 scheme using the tensor cores on an Nvidia A100 GPU. To take full advantage of the speed increase demonstrated in Figure~\ref{fig:sp2 vs cusolver}, we do not use the refinement step for the density matrix calculation. By comparing the tensor core evaluated forces with double-precision evaluated forces at each time step (as seen in Figure~\ref{fig:force-distribution}), the magnitude of the tensor core noise in the forces can be estimated. For the water box this is estimated to be $\sigma_{\rm TC} = 1.5\times 10^{-4}$ so that, at room temperature ($T = 300$ K) with $\delta t = 0.5$ fs, this leads to $\gamma_{\rm TC} = 2.3 \times 10^{-7}$ fs$^{-1}$ from Eq.\ (\ref{eq:fluct_diss}). Figure~\ref{fig:temp-distribution} shows the distribution of the temperature fluctuations over time generated from a 100 ps QMD simulation of the water system in comparison to the theoretically exact canonical temperature distribution, where temperatures were calculated using the statistically correct half-step velocity defined in Ref.~\citenum{NGronbechjensen19}. The integration scheme in \cref{IntegrationAlg}, with the fluctuation-dissipation in \cref{eq:fluct_diss}, is used to numerically integrate the nuclear degrees of freedom with a time step of $\delta t = 0.5$ fs and an additional regular Langevin friction constant of $\gamma_L = 10^{-3}$ fs$^{-1}$. The example used for Figure~\ref{fig:temp-distribution} demonstrates our ability to perform efficient and accurate canonical QMD simulations using tensor cores. There is virtually no difference between the histogram of the thermal fluctuations estimated from the QMD simulations and the theoretically exact distribution.

    \section{Conclusions}

    Tensor cores, along with tensor processing units, represent a new form of hardware acceleration that can deliver extraordinary computational speed and energy efficiency. However, they were principally designed for tensor contractions in machine learning applications using only low precision floating point operations. In spite of this, we have demonstrated how tensor cores can be applied, with high efficiency, to the challenging task of QMD simulations. This was achieved using a carefully tailored multidisciplinary approach. The interatomic forces were calculated from an electronic structure that was obtained from a generalized deep neural network. This neural network has a computational structure that can naturally harness the exceptional processing power of the tensor cores. The tensor contractions of the deep-neural network were performed in mixed-precision using dual matrix factors in half-precision representations, whose products were accumulated in single-precision. This doubles the cost, but reduces the numerical uncertainty and achieves close to single-precision accuracy. A performance in excess of 100 Tflops was demonstrated using the tensor cores of a single Nvidia A100 GPU. Stable molecular dynamics trajectories were then generated using the framework of XL-BOMD, which combines computational efficiency with long-term stability, even under approximate charge relaxations and force evaluations that are limited in accuracy due to the low precision floating-point operations. XL-BOMD not only improves stability under numerically noisy conditions, but it also removes the computational overhead when compared to regular direct BOMD by avoiding the iterative electronic ground state optimization. Further, by embedding the numerical error from the tensor core calculations into the interatomic forces as an additional noise term in a Langevin-like dynamics, we proposed and demonstrated an integration scheme for canonical simulations. This integration scheme works for general non-normally distributed random errors in the interatomic forces.

    The ability to perform reliable and stable QMD simulations opens up new avenues of application for tensor cores, and more generally, tensor processing units in chemistry, materials science and molecular biology. The current development mirrors a similar transition that started more than a decade ago when specialized GPUs gradually became accessible to more general scientific computations. 
    
    It is worth emphasizing that the straightforward approach of combining tensor cores to solve the electronic structure problem with a regular direct Born-Oppenheimer molecular dynamics would not work well. It was the careful reformulation and combination of methods presented in this paper that was necessary to achieve the demonstrated level of performance. Indeed even as hardware becomes more efficient and performant, we expect the same basic advantage to hold: low precision arithmetics on specialized hardware, such as tensor cores and tensor processing units~\cite{CYoung17,ZPan21}, or field programmable gate arrays~\cite{Yang19} and neuromorphic processors~\cite{Mniszewski19,Aimone18,MDavies18,XXu21}, will be substantially faster than higher precision arithmetics on those same architectures. In this way, our results present a general road-map for higher performance QMD simulations also using future accelerated hardware.
    
    \section{Acknowledgments}
    This work is supported by the U.S. Department of Energy Office of Basic Energy Sciences (FWP LANLE8AN), the LANL LDRD-ER program, and by the U.S. Department of Energy through the Los Alamos National Laboratory. We thank the CCS-7 group and Darwin cluster at Los Alamos National Laboratory for computational resources. Darwin is funded by the Computational Systems and Software Environments (CSSE) subprogram of LANL’s ASC program (NNSA/DOE). We are thankful to Nicolas Bock for his advice on code development. We acknowledge future fruitful contributions and stimulating discussions from Travis Peery and the Los Alamos T-Division Ten Bar Java Group.

    \bibliography{main.bib}

    \end{document}


\title{Supporting Information}

\maketitle

\section{Python implementation of integration \\ scheme with non-Gaussian noise}

\begin{figure} 
\centering 
\begin{minipage}{0.45\linewidth}
    \includegraphics[width=\textwidth]{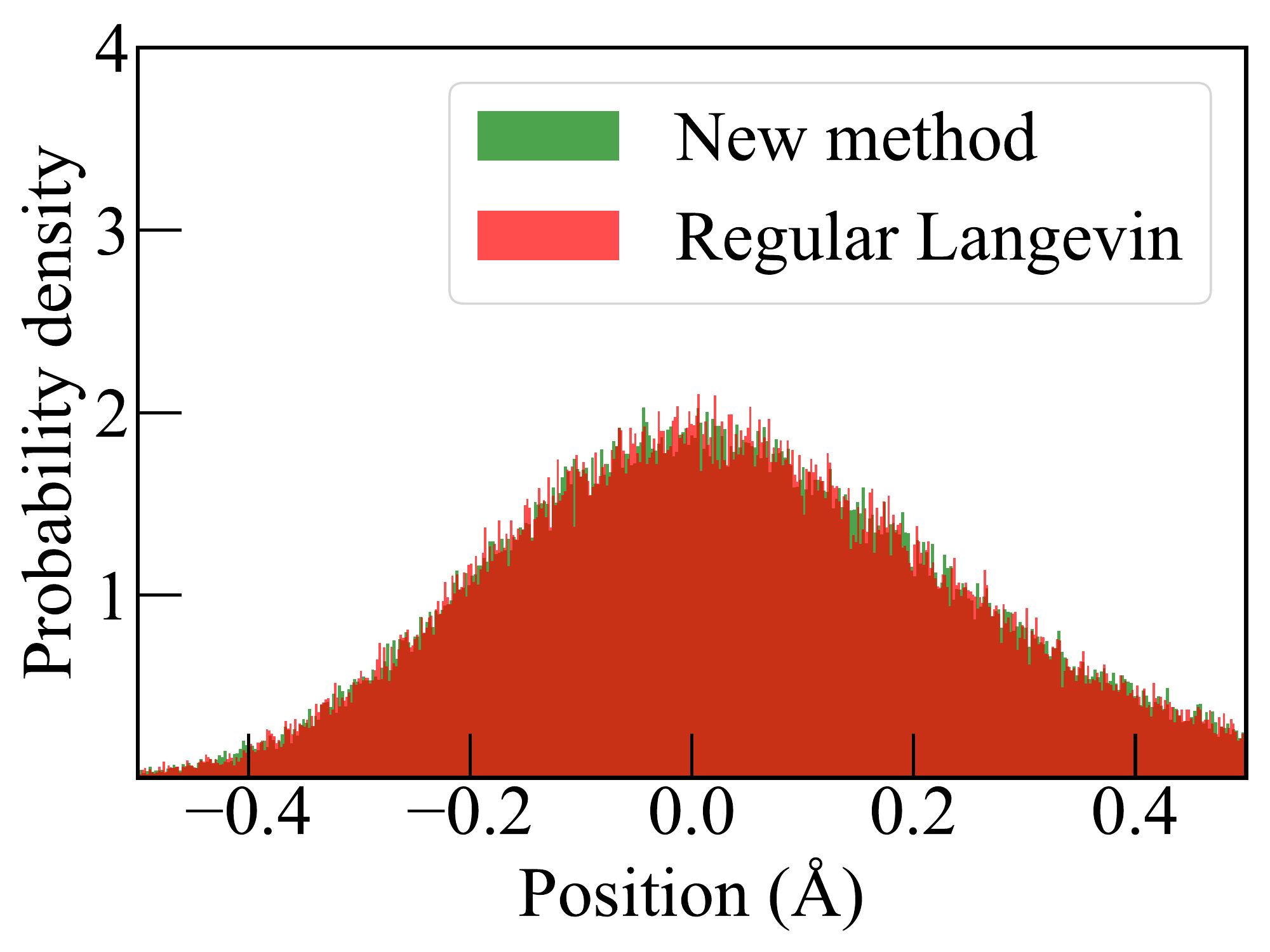} 
    \caption{GJF method compared to the canonical method presented in the main body of the text. Parameters used: $\sigma_{\rm TC} = 10^{-1}$ eV/\AA, $T = 500$ K, $M = 7$ amu, $\gamma_L = 0$ and $\delta t=0.5$ fs for a Morse potential.}\label{fig:main-scheme-histogram-1}
\end{minipage}
\hfill \begin{minipage}{0.45\linewidth}
    \vspace{.03in}
    \includegraphics[width=\textwidth]{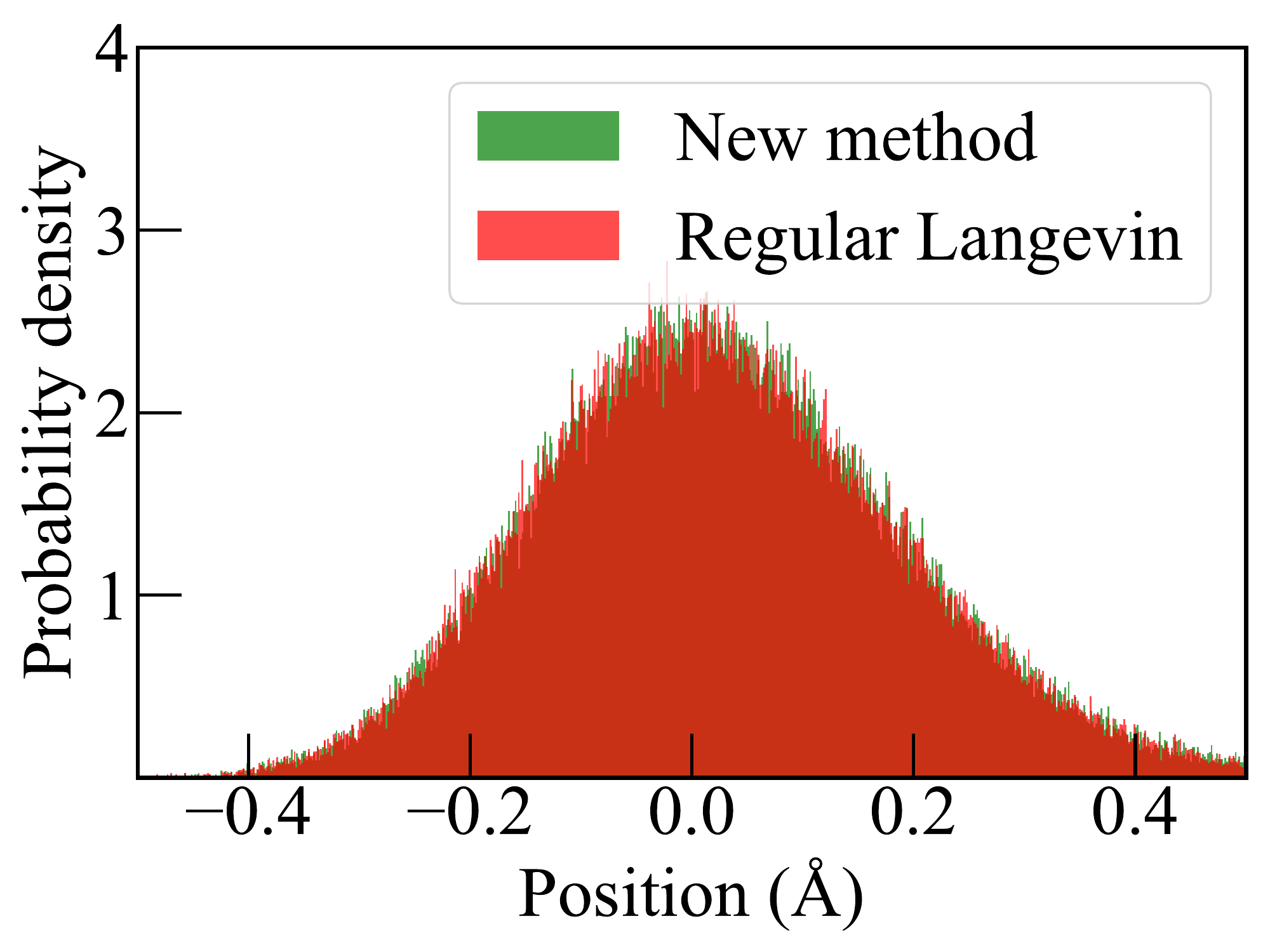} 
    \caption{GJF method compared to the canonical method presented in the main body of the text. Parameters used: $\sigma_{\rm TC} = 10^{-2}$ eV/\AA, $T = 300$ K, $M = 7$ amu, $\gamma_L = 0$ and $\delta t=0.5$ fs for a Morse potential.}\label{fig:main-scheme-histogram-2}
\end{minipage}
\end{figure}

The following Python script implements the time integration method in Eq. (31) with the only noise present coming from a non-normal random variable added to the force evaluation. A standard Langevin method, GJF, using a normally distributed noise term is simulated in parallel, so that a comparison of configurational distributions can be made between a standard Langevin dynamics and the scheme in Eq. (31) of the main text. Using a Morse potential, fully equilibrated histograms for different choices of simulation parameters are presented below in \cref{fig:main-scheme-histogram-1,fig:main-scheme-histogram-2}. To illustrate how the method works without additionally added-on noise, we set $\gamma_L = 0$.

\begin{lstlisting}
import numpy as np
import matplotlib.pyplot as plt
import matplotlib

def rand_rect(x):
  n = len(x)
  r = np.sqrt(12/17)*(4*np.random.uniform(0,1,num_atoms)+1 \
                 + (np.random.uniform(0,1,num_atoms) - 3.5))
  return r

if __name__ == "__main__":

  ##
  ## Initialize parameter choices
  ##
  dt = 0.5; M = 7.0; kb = 8.617e-5
  a=np.sqrt(2); T = 300.0; Etot=[]; Temp=[]
  num_steps = int(1E5); num_atoms = int(1E5)

  ##
  ## Initialize positions and velocities
  ##
  x = 0.50*(np.random.uniform(0,1,num_atoms)-.5)
  v = np.zeros(num_atoms); r = x; s = v

  ##
  ## Set seed
  ##
  np.random.seed(1)

  ##
  ## Tensor core noise parameters
  ##
  Sigma_TC = 0.01
  gamma_TC = (Sigma_TC)**2*dt/(2*kb*T)
  c_TC = (1-gamma_TC*dt/2)/(1+gamma_TC*dt/2)

  ##
  ## Langevin dynamics parameters
  ##
  gamma_L=gamma_TC
  c_L = (1-gamma_L*dt/2)/(1+gamma_L*dt/2)
  Sigma_L = np.sqrt(kb*T*(1-c_L**2)/M)

  ##
  ## Compute F_0
  ##
  F_L = -a*np.exp(-x/a)*(1-np.exp(-x/a))
  TC_Force_Noise = Sigma_TC*rand_rect(r)
  F_TC = -a*np.exp(-r/a)*(1-np.exp(-r/a)) \
       + TC_Force_Noise 

  ##
  ## Main MD loop
  ##
  for j in range(0,num_steps):

    ##
    ## Compute V_{n+1/4}
    ##
    s = s + 0.5 * dt * (F_TC - gamma_TC*s) / M
    v = v + 0.5 * dt * F_L / M

    ##
    ## Compute R_{n+1/2}
    ##
    r = r + 0.5 * dt * s
    x = x + 0.5 * dt * v

    ##
    ## Compute V_{n+3/4}
    ##
    v = v * c_L + Sigma_L * np.random.normal(0,1,len(s))

    ##
    ## Compute R_{n+1}
    ##
    r = r + 0.5 * dt * s
    x = x + 0.5 * dt * v

    ##
    ## Compute F_{n+1}
    ##
    TC_Force_Noise = Sigma_TC*rand_rect(r)
    F_TC = -a*np.exp(-r/a)*(1-np.exp(-r/a)) + TC_Force_Noise
    F_L = -a*np.exp(-x/a)*(1-np.exp(-x/a))

    ##
    ## Compute V_{n+1}
    ##
    s = (s + 0.5 * dt * F_TC / M) / (1 + 0.5 * dt * gamma_TC / M)
    v = v + 0.5 * dt * F_L / M

    ##
    ## Record temperature each step
    ##
    EKin = 0.5*M*sum(s**2)
    Temp.append((2*EKin/kb)/len(r))


  ##
  ## Plotting section
  ##
  plt.rcParams["font.family"] = "Times New Roman"
  matplotlib.rc('axes',
                edgecolor='black',
                linewidth=2.0)

  num_bins=1000
  
  plt.hist(r, num_bins,
           density = 1,
           color = 'green',
           alpha = 0.7,
           label = 'New method')
  plt.hist(x, num_bins,
           density = 1,
           color = 'red',
           alpha = 0.7,
           label = 'Regular Langevin')

  plt.xlabel('Position (\305)',
             fontsize=20)
  plt.ylabel('Probability density',
             fontsize=20)
  plt.xticks(np.arange(-0.6,0.7, step =0.2),
             fontsize=20)
  plt.yticks(np.arange(1,3,step=1.),
             fontsize=20)
  plt.xlim([-5e-1,5e-1])
  plt.legend(fontsize=20)
  plt.tick_params(which='major',
                  axis='both', 
                  direction='in',
                  length=20.0,
                  width=1.5)
  plt.tight_layout()
  plt.show()
\end{lstlisting}

\newpage

\section{Alternative integration scheme}

\begin{figure} 
\centering 
\begin{minipage}{0.45\linewidth}
    \includegraphics[width=\textwidth]{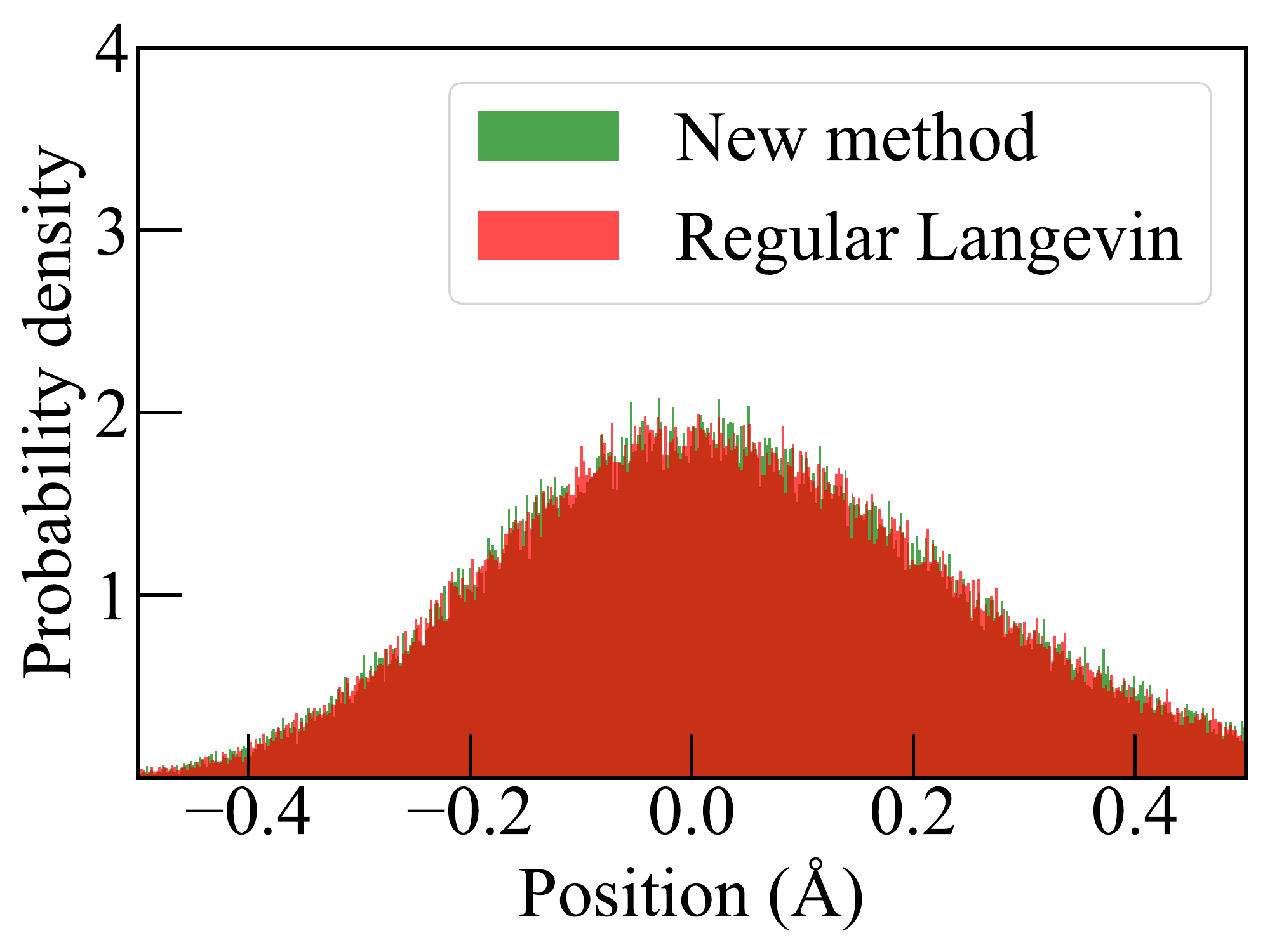} 
    \caption{GJF method compared to the method presented in \cref{IntegrationAlg_New} of the Supporting Information. Parameters used: $\sigma_{\rm TC} = 10^{-1}$ eV/\AA, $T = 500$ K, $M = 7$ amu, $\gamma_L = 0$ and $\delta t=0.5$ fs for a Morse potential.} \label{fig:new-scheme-histograms-3}
\end{minipage}
\hfill \begin{minipage}{0.45\linewidth}
    \vspace{.03in}
    \includegraphics[width=\textwidth]{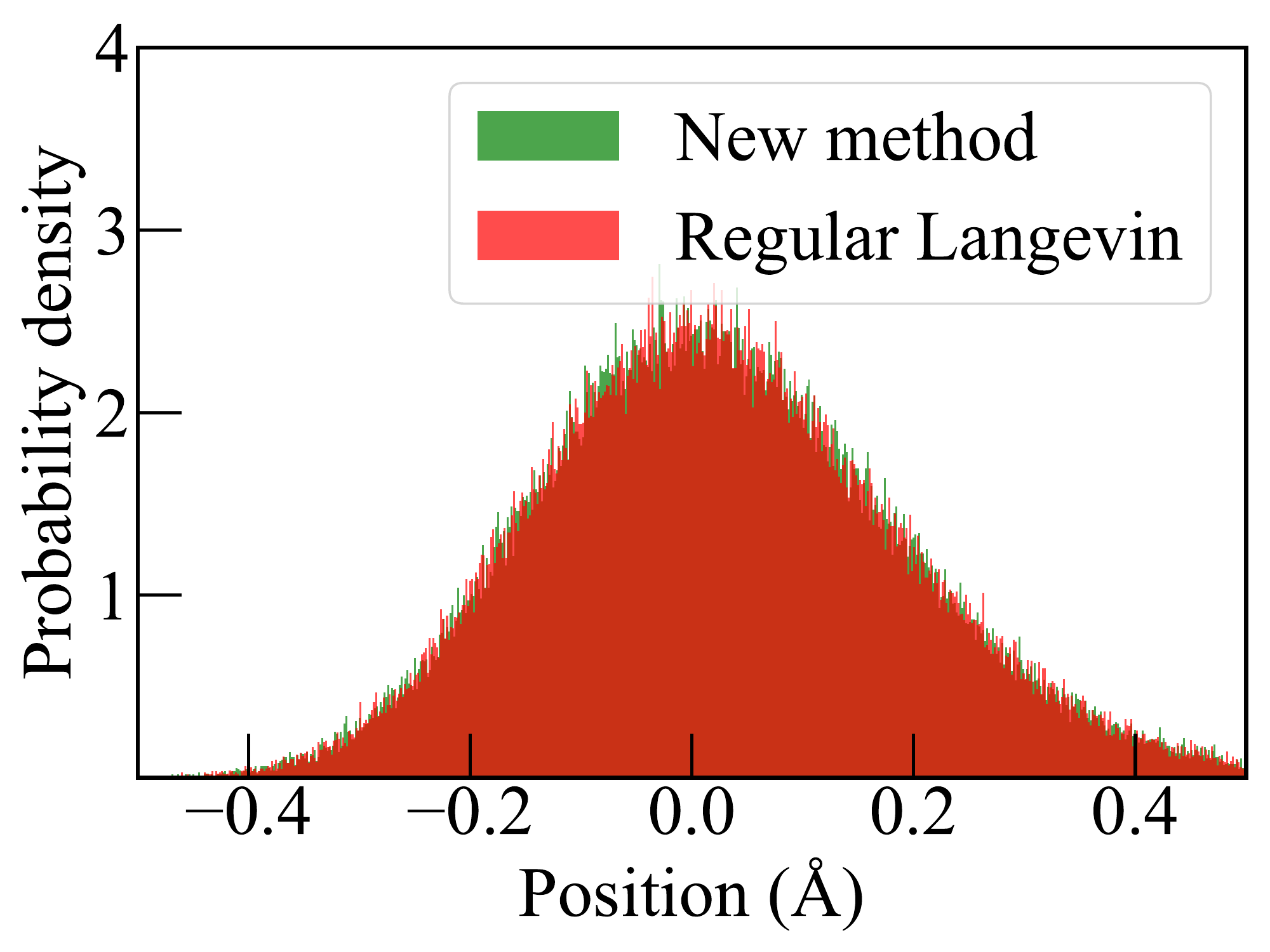} 
    \caption{GJF method compared to the method presented in \cref{IntegrationAlg_New} of the Supporting Information. Parameters used: $\sigma_{\rm TC} = 10^{-2}$ eV/\AA, $T = 300$ K, $M = 7$ amu, $\gamma_L = 0$ and $\delta t=0.5$ fs for a Morse potential. } \label{fig:new-scheme-histograms-4}
\end{minipage}
\end{figure}

In Eq. (31) we presented only one of many possible integration schemes which utilizes the tensor core noise as a heat bath. Here we present an alternative method that can be constructed also using only a single tensor core force evaluation per time step. Fully equilibrated histograms for a Morse potential and different simulation parameters are presented below in \cref{fig:new-scheme-histograms-3,fig:new-scheme-histograms-4}.

    \begin{align}\label{IntegrationAlg_New}
        \begin{split}
        {V}_{k+1/5} &= {V}_k + \frac{\delta t}{2M} {F}_k^{\rm TC}  \\
        {R}_{k+1/2} &= {R}_k + \frac{\delta t}{2} {V}_{k+1/5}\\
        {V}_{k+2/5} &= \sqrt{c_{\rm TC}} \; {V}_{k+1/5} \\
        {V}_{k+3/5} &= c_L \; {V}_{k+2/5} + {\sigma}_L \; {\eta}_k\\
        {V}_{k+4/5} &= \sqrt{c_{\rm TC}} \; {V}_{k+3/5} \\
        {R}_{k+1} &= {R}_{k + 1/2} + \frac{\delta t}{2} {V}_{k+4/5}\\
        {V}_{k+1} &=  V_{k+4/5} + \frac{\delta t}{2M}  F_{k+1}^{\rm TC} \;,
        \end{split}
    \end{align}
which has discrete-time fluctuation dissipation relation
\begin{align}
    \sigma_{\rm TC}^2 = \frac{2k_{\rm B} T \gamma_{\rm TC} M}{\delta t} \;,
\end{align}
and $c_{\rm TC}$ chosen so that 
\begin{align}
    c_{\rm TC} = \frac{1-\gamma_{\rm TC} \delta t/2}{1+\gamma_{\rm TC} \delta t/2} \;.
\end{align}
The main structural difference between the previously proposed method in Eq. (31) of the main text and \cref{IntegrationAlg_New} above is that the dissipation is now included mid-step by itself, instead of in the same step as the noise. This formulation is more in-line with the BAOAB \cite{BLeimkuhler13} and GJF \cite{NGronbechjensen13} Langevin equation time integration methods, which have been shown to have optimal statistical properties for linear potentials. 

\newpage
\bibliography{si}